\documentclass{aa}  
\usepackage{graphicx}
\usepackage{txfonts}
\usepackage{xcolor}

\newcommand{\Teff}{\ensuremath{T_\mathrm{eff}}}

\newcommand{\feh}{[Fe/H]}

\newcommand{\meanks}{$\langle K_S \rangle$}

\def\deg{${}^\circ$}

\begin{document}

   \title{Using classical Cepheids to study the far side of the Milky Way disk \thanks{Based on observations collected at the European Southern Observatory under ESO programmes 095.B-0444(A), 179.B-2002}}
   \subtitle{II. The spiral structure in the first and fourth Galactic quadrants}
   
   \author{J.~H.~Minniti\inst{1,2,3}
   \and
   M.~Zoccali\inst{1,2}
   \and
   A.~Rojas-Arriagada\inst{1,2}
   \and
   D.~Minniti\inst{2,4,5}
   \and
   L.~Sbordone\inst{3}
   \and
   R.~Contreras Ramos\inst{1,2}
   \and
   V.~F.~Braga\inst{2,4}
   \and
   M.~Catelan\inst{1,2}
   \and
   S.~Duffau\inst{4}
   \and
   W.~Gieren\inst{2,6}
   \and
   M.~Marconi\inst{7}
   \and
   A.~A.~R.~Valcarce\inst{8,9,10}
   }
   \institute{Pontificia Universidad Cat{\'o}lica de Chile, Instituto de Astrof{\'i}sica, Av. Vicu\~na Mackenna 4860, 7820436, Macul, Santiago, Chile\\
              \email{jhminniti@uc.cl}
         \and 
             Millennium Institute of Astrophysics, Av. Vicu\~na Mackenna 4860, 7820436, Macul, Santiago, Chile
         \and
             European Southern Observatory, Alonso de C{\'o}rdova 3107, Santiago, Chile
         \and
             Departamento de F{\'i}sica, Facultad de Ciencias Exactas, Universidad Andr{\'e}s Bello, Fern{\'a}ndez Concha 700, Las Condes, Santiago, Chile
          \and
             Vatican Observatory, V00120 Vatican City State, Italy
           \and
             Universidad de Concepción, Departamento de Astronomía, Casilla 160-C, Concepción, Chile
          \and
             INAF-Osservatorio Astronomico di Capodimonte, Via Moiariello 16, 80131 Naples, Italy
         \and
             Departamento de F{\'i}sica, Universidade Estadual de Feira de Santana, Av. Transnordestina, S/N, CEP 44036-900 Feira de Santana, BA, Brazil
          \and
             Centro de Nanotecnolog{\'i}a Aplicada, Facultad de Ciencias, Universidad Mayor, Santiago, Chile
          \and
             Centro para el Desarrollo de la Nanociencia y la Nanotecnolog{\'i}a (CEDENNA), Santiago, Chile
             }

   \date{Received 23 September 2020 / Accepted}

 
  \abstract
   {In an effort to improve our understanding of the spiral arm structure of the Milky Way, we use Classical Cepheids (CCs) to increase the number of young tracers on the far side of the Galactic disk with accurately determined distances. We use a sample of 30 CCs, discovered using near-infrared photometry from the VISTA Variables in the Vía Láctea survey (VVV) and classified based on their radial velocities and metallicities. We combine them with another 20 CCs from the literature for which VVV photometry is available. The compiled sample of CCs with homogeneously computed distances based on VVV infrared photometry was employed as a proof of concept to trace the spiral structure in the poorly explored far side of the disk. Although the use of CCs has some caveats, these variables are currently the only available young tracers in the far side disk for which a numerous sample with accurate distances can be obtained. Therefore, a larger sample could allow us to make a significant step forward in our understanding of the Milky Way disk as a whole. We present preliminary evidence that CCs favor: a spiral arm model with two main arms (Perseus and Scutum-Centaurus) branching out into four arms at galactocentric distances, $R_\mathrm {GC}\gtrsim5-6\,\mathrm{kpc}$; the extension of the Scutum-Centaurus arm behind the Galactic center; a possible connection between the Perseus arm and the Norma tangency direction. The current sample of CCs in the far side of the Galaxy are in the mid-plane, arguing against the presence of a severely warped disk at small Galactocentric distances ($R_\mathrm {GC}\lesssim12\,\mathrm{kpc}$) in the studied area. The discovery and characterization of CCs at near-IR wavelengths appears to be a promising tool to complement studies based on other spiral arm tracers and extend them to the far side of our Galaxy.}


   \maketitle


\section{Introduction}

The complete picture of our Galaxy, the Milky Way (MW), is still to be unveiled. In particular, our knowledge of the Galactic spiral structure is still far from complete, due to its intrinsic complexity together with our edge-on view of the disk and location at $\sim8.2\,\rm{kpc}$ from the center \citep{blandhawthorn16}.

Star formation in the MW disk happens predominantly along the spiral arms. The youngest stellar populations did not migrate significantly from their birthplace, so that the position of spiral arms can be mapped by tracing the position of massive young stars, as well as of neutral hydrogen, molecular clouds or \ion{H}{ii} regions associated with high-mass star formation \citep[see][]{vallee2017b}. One advantage of using these latter tracers is that they are observed in the radio regime, making them largely insensitive to extinction. Even in these cases, however, precise distances through Very Long Baseline Interferometry (VLBI) parallaxes are scarce on the far side of the disk, while most are determined kinematically and have large errors that prevent to delineate the spiral arm structure \citep[see][for a recent review]{xu2018}. Regardless of the obstacles, our knowledge of the spiral structure of the MW has experienced a significant improvement over the past decades. A clear picture of the local, near side, spiral arm pattern is emerging from both VLBI distance measurements of masers associated with high-mass star forming regions \citep[HMSFRs, see][and references therein]{reid2014,reid2019} and OB stars with accurate {\em Gaia} parallaxes \citep{chen2019}, which show consistent spatial distributions \citep[][]{xu2018a,xu2018}.

Maser parallaxes from VLBI observations have provided a detailed and accurate view of Galactic spiral arm structure. The main parameters for each arm (pitch angle, reference radius, arm-width) have been obtained from these observations \citep[][]{reid2014, reid2019, xu2016}. This allowed to trace arm segments in a region covering around one fourth of the disk extension, that have been interpreted as corresponding to 4 major spiral arms. So far, this method has been limited to the portion of the disk observable from the northern hemisphere and belonging to the near side of the MW disk (the only exception being the work of \citeauthor{sanna2017} \citeyear{sanna2017}, who mapped the passage of the Scutum-Centaurus arm through the far side of the disk based on the parallax measurement for a maser at $\sim20\,\mathrm{kpc}$), leaving the far side of the first and the whole fourth Galactic quadrant almost devoid of young tracers with accurate distance determinations.

High dust extinction in the optical limits the use of young stellar tracers in combination with {\em Gaia} parallaxes to the near side of the disk, reaching distances up to $\sim6\,\mathrm{kpc}$ from the Sun using luminous OB stars \citep{xu2018a,chen2019}. Infrared (IR) observations are needed to thoroughly explore the heavily obscured Galactic midplane.

Classical Cepheids (CCs) are pulsating variable stars known to follow accurate period-luminosity (PL) relations (particularly in the near-IR) from which we can determine accurate distances. They are key to calibrate the cosmic distance ladder \citep{ripepi2020}. Due to their high luminosities and young ages, CCs are ideal objects to map the structure of the thin disk down to the edge of the Galaxy, and they have been used to construct a map of the warp of the disk \citep[see][]{skowron2019,skowron2019b,dekany2019}. These structural studies can also be done in nearby spiral \citep[e.g., M31,][]{kodric2018}, and dwarf irregular galaxies \citep[][for the Large and Small Magellanic Clouds]{inno2016,ripepi2017}. CCs have also been useful to trace the metallicity gradient of the Galactic disk \citep[see][and references therein]{genovali2014,luck2018, lemasle2018}. In particular, if observed in the near-IR, the aforementioned extinction issue can be mitigated and CCs could be used to map the portion of the Galactic disk otherwise hidden by dust \citep[see also][]{dekany2019}.

In \citet[][hereafter Paper I]{minniti2020} we analyzed a sample of 45 candidate Cepheids detected in the VISTA Variables in the Vía Láctea \citep[VVV,][]{minniti2010} ESO Public Survey and located within $|b| < 1.5$\degr. By means of their radial velocities and metallicities, we isolated a clean sample of 30 CCs lying on the far side of the disk. We used these CCs to derive an extinction law for the MW midplane and a disk metallicity gradient in the range $5\,\mathrm{kpc}\lesssim R_\mathrm{GC} \lesssim 22\,\mathrm{kpc}$. Here we use the sample of 30 CCs, complemented with another 20 detected and classified from both VVV and the Optical Gravitational Lensing Experiment (OGLE) surveys, to trace the far side disk spiral structure.

\section{A bona fide sample of classical Cepheids on the far side of the Galactic disk}

Thanks to their characteristic pulsation (light-curves), CCs are relatively easy to identify, spanning a range of periods of $\sim \rm{1-100}\,\rm{days}$ \citep[][]{catelan2015}. This is particularly true in the optical, where more than two thousand Galactic CCs have been catalogued so far, mainly by recent time-domain surveys \citep[e.g., the OGLE Collection of Galactic Cepheids; see][]{udalski2018}. When moving to the near-IR bands, the classification of pulsating stars becomes harder, since their light curves have smaller amplitudes and become more sinusoidal. Moreover, near-IR observations are more time-consuming, and time domain surveys at those wavelengths have generally fewer epochs than their optical counterparts. Nevertheless, IR variability searches are the only solution to detect most CCs at the far side of the MW disk, where optical surveys (such as OGLE and {\em Gaia}) are far from complete due to the extreme interstellar dust extinction towards the midplane (see section~\ref{subsection:R-Z_plane}).

The CCs used here were classified in Paper I, from spectra taken with the X-Shooter spectrograph \citep{vernet2011}, located at the ESO Very Large Telescope, of a larger number of candidate Cepheids selected from VVV photometry. In order to be classified as CC, a Cepheid needed to fulfill the following criteria: i) the Cepheid radial velocity was within $\approx30\,\mathrm{km\, s^{-1}}$ from the expected value for the Galactic disk if it were to follow the circular rotation at its corresponding position; ii) its \feh{} was $\gtrsim -0.6\,\mathrm{dex}$; iii) its \Teff{} was consistent with its CC nature. This left us with 30 CCs spanning $+10\degr>\ell>-40\degr$, ages mainly between 40 and 85\,Myr \citep[][ their theoretical period-age relations obtained for a non-canonical mass-luminosity relation]{desomma2020}, and galactocentric distances $5\,\mathrm{kpc}<R_\mathrm{GC}<31\,\mathrm{kpc}$, calculated based on the PL relations by \cite{macri2015} and the extinction law derived in Paper I. For the details about the distance determination, the reddening law and the photometry for each CC, please refer to Paper I\footnote{We remind the reader that the photometric issues discussed in \cite{hajdu2019} were taken into account, as described in Paper I.}.

With the purpose of increasing the size of the sample of CCs on the far side of the disk with reliable distances obtained from near-IR photometry, the stars classified by the OGLE collection of Galactic Cepheids \citep{udalski2018} that were also present in the VVV footprint were included here. We retrieved their VVV photometry from the catalogue presented in \cite{dekany2019}. We also incorporated a subsample of Cepheid candidates classified by \cite{dekany2019} that were recently confirmed as CCs based on OGLE photometry by \cite{soszynski2020}\footnote{We only use CCs from the sample by \cite{dekany2019} that were confirmed by OGLE. There is evidence that their classification, based on the near-IR light curve, suffers from a large contamination; see \url{http://www.astrouw.edu.pl/ogle/ogle4/OCVS/Cepheid_Misclassifications/Dekany_et_al_VVV/}}. Before using them we decided to select stars with \meanks$\,\gtrsim 11\,\rm{mag}$ to avoid light curves with heavily saturated VVV photometry. Moreover, these CCs have, in general, smaller periods than those from Paper I, and are thus older. We selected stars with $\rm{P}>5\,\rm{days}$ as a trade-off between keeping as many stars as possible to increase our sample, while avoiding the lower period and thus older CCs. As a result, we added 20 CCs with ages between 50 and 110\,Myr, that are also complementary in terms of their distribution in the Galaxy, as we will show later. It is important to emphasize that these cuts are somewhat arbitrary and could be modified in further studies, as the sample of well-known CCs in the far side increases and allows to study their distribution as a function of age. The sample of 50 bona fide CCs with homogeneously determined distances based on VVV near-IR photometry can be used to study the spiral structure of the far disk.

\section{The spiral structure in the first and fourth Galactic quadrants}

\begin{figure}
\centering
    \includegraphics[width=8.5cm]{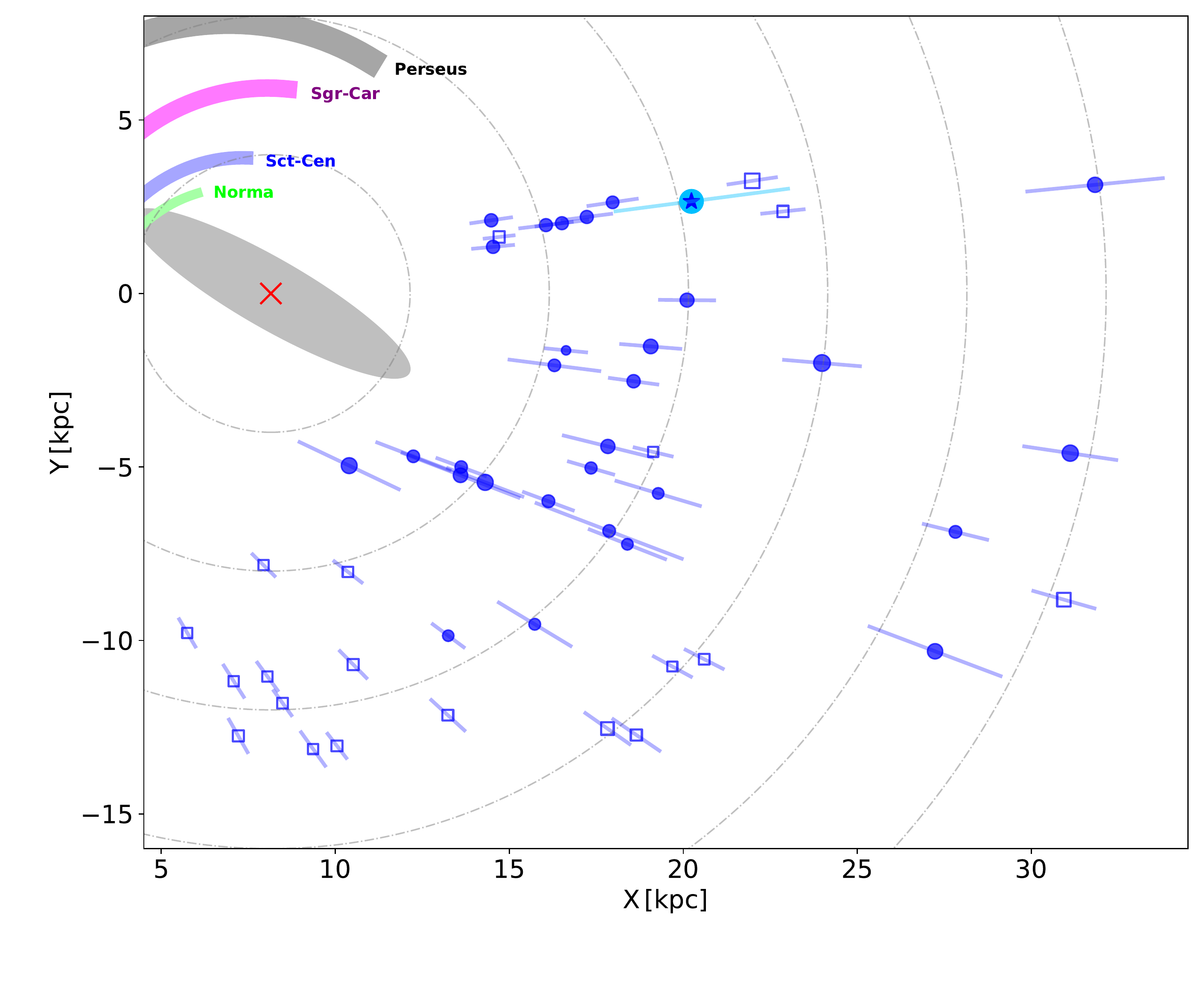}
       \caption{In-plane distribution of the sample of 50 far disk bona fide CCs with homogeneously determined distances based on VVV near-IR photometry. Our sample of spectroscopically confirmed CCs is shown here using blue circles. The blue open squares represent the complementary sample of OGLE-VVV CCs. Distance error bars are indicated for each star and the point sizes are inversely proportional to the Cepheid ages. The far side disk maser source presented in \cite{sanna2017} (blue star inside circle) is shown with its corresponding error bar. The Sun is located at ($X=0\,\mathrm{kpc}$, $Y=0\,\mathrm{kpc}$), and the Galactic center is marked with a red cross at ($X=8.15\,\mathrm{kpc}$, $Y=0\,\mathrm{kpc}$). The colored sections are the log-periodic spiral arm segment fits based on VLBI trigonometric parallaxes of HMSFRs by R19. The dash-dotted circles in the background mark galactocentric radii ranging from 4 to $24\,\mathrm{kpc}$.}
       \label{fig:xyplane_errorbars}%
\end{figure}

With our 30 spectroscopically classified CCs together with other 20 Cepheids classified based on optical light-curves, we analyse their projection on the far side of the Galactic plane (see Fig.~\ref{fig:xyplane_errorbars}). In order to investigate the spiral structure by means of CCs, we need to make some assumptions about the relative velocity of the arms with respect to the variables, and the stars in general. Indeed, because the ages of CCs are not zero, and given that the spiral arm pattern may move with respect to the stars, the current position of the CCs might not trace the arms, if their individual rotation velocities differ significantly from the velocity of the spiral pattern.

Because the spiral pattern speed is still poorly constrained, as it is the fact that spiral arms move as a rigid-body, we follow here two approaches, explained below. We refer the reader to \cite{dobbs&baba2014} for a review of the different mechanisms proposed for the formation of the spiral arms, and the consequent expected rotation of their pattern.

First, in subsection~\ref{subsection:xyplane}, we show the current position of the CCs, and compare them with the current position of the spiral arms. This approach makes sense if the spiral arm pattern has a similar rotation speed to that of rotation of the stars. Although this might not be a perfect scenario (though it is not discarded either), it is the simplest hypothesis, that allows us to show the data with minimal corrections. This is useful for future analysis, when our understanding of spiral arms formation and kinematics improves. In addition, as we will see, it shows a very good agreement between the expected (extrapolated) location of the arms, based on how they were traced in the disk's near side, and the current position of CCs. Something similar has been found in M31 \citep[see][]{kodric2018}, where the actual position of CCs correlates tightly with the position of the star-forming ring and the structures traced by dust (their Fig 17). In subsection~\ref{subsection:patternspeed} instead, we will assume that spiral arms move as a rigid-body with a rotation period of 250\,Myr, while CCs (like other stars) follow the disk rotation curve. We will therefore adjust the position of the CCs by the difference in their velocity with respect to the arms during their lifetimes, and compare the two. Finally, in subsection~\ref{subsection:R-Z_plane} we discuss the vertical distribution of our CCs.

\subsection{In-plane distribution}\label{subsection:xyplane}

\begin{figure*}
\centering
    \includegraphics[width=17cm]{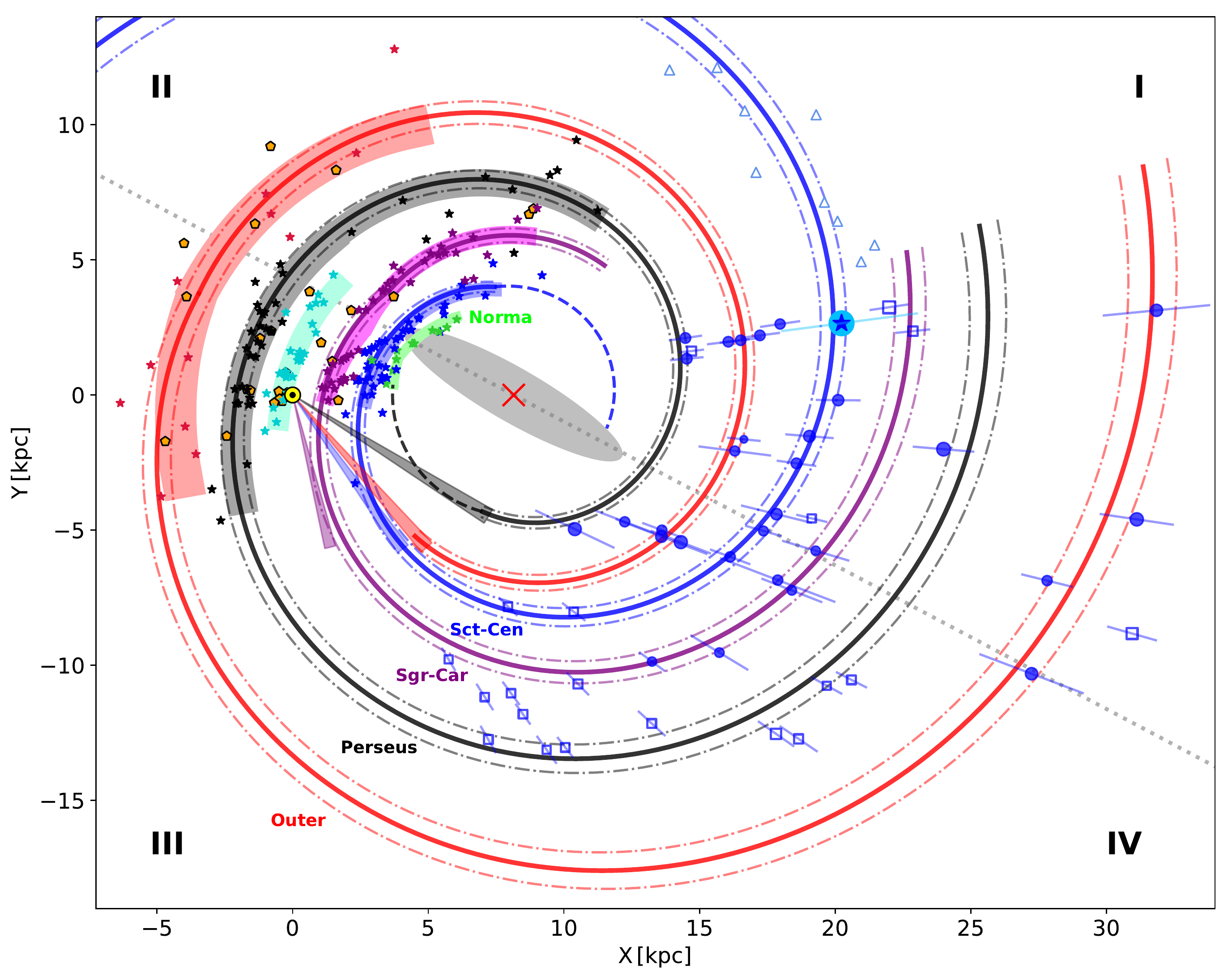}
       \caption{Milky Way spiral structure as traced by CCs and HMSFRs. The distribution of our sample of spectroscopically confirmed CCs in the first and fourth quadrants is shown here (blue circles). We overplot the OGLE CCs with distances based on near-IR photometry (blue open squares). Distance error bars are indicated for each star. The point sizes are inversely proportional to the Cepheid ages. We also included the maser source presented in \cite{sanna2017} (blue star inside circle). The Sun is located at ($X=0\,\mathrm{kpc}$, $Y=0\,\mathrm{kpc}$) and marked with a yellow dot, and the Galactic center is marked with a red cross at ($X=8.15\,\mathrm{kpc}$, $Y=0\,\mathrm{kpc}$). As in Fig.~\ref{fig:xyplane_errorbars}, the log-periodic spiral arm segment fits based on VLBI trigonometric parallaxes of HMSFRs by R19 are represented with colored sections. The turquoise spiral segment is the Local arm. The masers associated with each arm in R19 are overplotted using stars (Norma arm, green; Sct-Cen arm, blue; Sgr-Car arm, purple; Local arm, turquoise; Perseus arm, black; Outer arm, red). The thick lines represent our qualitative logarithmic spiral arm fits, with slight changes to R19 best-fit coefficients in order to have them passing through our CCs, with the thin dashed-lines marking the $1\sigma$ arm widths reported in R19. We also included maser sources associated with HMSFRs recently reported in the VLBI Exploration of Radio Astrometry \citep[VERA;][]{vera2020} catalog (orange pentagons). The Carina, Centaurus (2 peaks), and Norma tangency directions are displayed as colored light cones originating from the Sun, using the same colors as for their associated spiral arms. The line of nodes of the Galactic warp model by \cite{skowron2019} is also shown for reference (gray dotted line). We have included the positions of the CO detections reported by \cite{dame&thaddeus2011} that trace the continuation of the Sct-Cen arm in the far side of the Galaxy (open triangles). Their kinematic distances were recalculated using the MW rotation curve determined by \cite{mroz2019}.}
       \label{fig:xyplane_last}%
\end{figure*}

We present here the current position of CCs on the far side of the Galactic plane and analyse how they fit into the present knowledge of the spiral arm structure. Given the large area that we are covering (around one fourth of the Galactic disk), the number of CCs utilized here is not sufficient to independently trace the position of the spiral arms. Therefore, as a first step, we verified their compatibility with the spiral arms as traced in the near side of the disk. We have used the results presented in \citet[][hereafter R19]{reid2019}, based on trigonometric parallaxes for masers obtained from VLBI. They fitted the HMSFRs positions using a log-periodic spiral functional form, allowing for a ‘‘kink’’ in the arms (the pitch angle, $\psi$, was allowed to change at this point), and provided the locations of several spiral arm segments which we plotted as colored sections in Fig.~\ref{fig:xyplane_last}. We only show them where they are constrained by VLBI data. If we extrapolate these segments as they are modelled in R19, both their Perseus and Scutum-Centaurus (Sct-Cen) best fit models are roughly consistent with our objects, while the Sagittarius–Carina (Sgr-Car) and Norma-Outer arms are not. However, a slight modification of the coefficients of the parametrization of the arms makes them qualitatively consistent with the position of our young stellar tracers, while they do not deviate significantly from the original formulation in the region where the arms were constrained by R19. We would like to emphasize here that the underlying assumption for the analysis presented in this subsection is that these stars were born in spiral arms and stayed close to their birthplaces during their lifetimes, which might be a good approximation just for the youngest CCs.

We derived the simplest possible parametrization of the spiral arms compatible with both the maser and CC positions, keeping a unique pitch angle over the full extent of the arm whenever possible. The log-periodic spiral form is defined as
\begin{equation} \label{eq:1}
\mathrm{ln} \left(R/R_{\mathrm{ref}}\right)  = - \left(\beta-\beta_{\mathrm{ref}}\right) \mathrm{tan}\,\psi,
\end{equation} where $R$ and $\beta$ are the Galactocentric radius and azimuth (with its origin toward the Sun and increasing clockwise), respectively. A more detailed explanation of the spiral model and its parameters can be found in Appendix~\ref{ap:log_per_spiral}.

Given that longer period CCs are younger, and thus expected to be closer to their birthplaces, we have given them preference while doing this qualitative comparison. Specifically, we gave preference to our spectroscopic sample of CCs, since they are mostly younger than the OGLE sample ($\lesssim85\,\rm{Myr}$) and to stars with uncertainties in their distances lower than $\sim 1.5\,\rm{kpc}$. In Figs.~\ref{fig:xyplane_errorbars} and~\ref{fig:xyplane_last} we assigned CCs point sizes inversely proportional to their ages. To calculate their Galactocentric distances, $R_{\mathrm{GC}}$, we assumed a Sun-Galactic center distance $R_0=8.15\,\mathrm{kpc}$, to simplify the comparison of their distribution with the spiral arm model of R19. The difference with the distance obtained by \cite{gravity2019} of $R_0=8178\pm13_{\rm{(stat)}}\pm22_{\rm{(sys)}}\,\rm{pc}$ is within the error bars.

Individual distance errors for the compiled sample of CCs are indicated with error bars in Fig.~\ref{fig:xyplane_errorbars}. They were calculated using Monte Carlo simulations, where the uncertainties on the mean J, H, and Ks magnitudes were considered, together with the intrinsic dispersion of the near-IR PL relations used \citep{macri2015} and their uncertainties, and the errors on the calculated selective-to-total extinction ratios, $A_{K_S}/E(H-K_S)$ and $A_{K_S}/E(J-K_S)$, obtained in Paper I. The distance uncertainties are between $\sim0.55\,\rm{kpc}$ and $\sim2.2\,\rm{kpc}$, with $\sim70\%$ having errors smaller than $1\,\rm{kpc}$ and $\sim90\%$ smaller than $1.4\,\rm{kpc}$. Although most of the individual error bars are larger than the arm widths, they are in general, smaller than the separation between the different arms, allowing in most cases to associate them with a unique spiral arm. As we have mentioned, the only object available in this region of the MW with a good distance measurement is the HMSFR by \cite{sanna2017}, with a distance error of $\sim2.5\,\rm{kpc}$. As can be noted in this Figure, our sample provides an independent improvement because most of our CCs have smaller distance errors than this unique maser source.

The parameters of the spiral arm model shown in Fig.~\ref{fig:xyplane_last} are presented in Table~\ref{tab:Spiral_arm_param}. Below, we discuss how the new constraints on the MW spiral structure obtained from CCs fit into its current picture.

\begin{table}
\footnotesize
\caption{Spiral Arm Characteristics.}
\centering
\begin{tabular}{lcccc}
\hline \hline
 Arm	    & $\beta_{\mathrm{ref}}$  &	$R_{\mathrm{ref}}$  &     $\psi$	&   $\ell$ tangency (Q4)    \\
            &  [\degr]       &       kpc   &	 [\degr]	&		[\degr]        \\
\hline
Sct-Cen	                 &	23   &	5.10   &	12.90       &   305.5, 311.2\tablefootmark{b}   \\
Sgr-Car                  &	40	 &  6.00   &    12.70       &   283.8   \\
Perseus                  &	40	 &  9.10   &    9.45        &   328.1\tablefootmark{*}   \\
Outer	                 &	18   & 12.20   &    10.50, 7.40 \tablefootmark{a}  &   --   \\
\hline
\end{tabular}
\tablefoot{\tablefoottext{a}{For this arm we allowed for a kink with pitch angles 10.5 and 7.4 for azimuths $\beta \leq \beta_{\mathrm{ref}}$ and $\beta > \beta_{\mathrm{ref}}$, respectively.}\tablefoottext{b}{Two peaks reported in \cite{hou+han2015}. The one at 305.5\degr is associated with the Sct-Cen arm, while for the second peak at 311.2\degr our data supports it is linked to the Outer arm.}\tablefoottext{*}{We propose the Norma tangency direction to be associated with the Perseus arm.}}
 \label{tab:Spiral_arm_param}
\end{table}

\paragraph{Perseus arm}We can trace the Perseus arm as it spirals inwards, behind the Galactic bulge, in the first and fourth quadrants, and also towards negative azimuthal angles, extending into the outer Galaxy. We find that a pitch angle of $9.45\degr$ is both consistent with the HMSFRs and the distribution of our CC sample at the far disk. This value is also in agreement with the pitch angle determined in \cite{zhang2019} ($5\degr\pm 4\degr$) for the inner portion of the Perseus arm and previous estimates of $\sim9\degr$, determined using sources along the whole spiral segment of this arm that has been traced based on HMSFRs distances \citep{reid2014,reid2019}. The proposed spiral arm parameters suggest that the Norma tangency point, usually used to constrain the shape of the Norma arm in the fourth Galactic quadrant and to connect it with the Outer arm, is instead associated with the Perseus arm. Although it is generally proposed in the literature that the Norma and Outer arm segments are connected, to the best of our knowledge there is no unequivocal evidence favouring this scenario. In R19, for example, this association was based on the inconsistency with the determined pitch angles of the other arms, but this argument depends on uncertain extrapolations over large sections of the disk. We are not in position of firmly establishing whether the Perseus and Norma arms are connected. Nonetheless, the continuation of the Perseus arm as it is modelled here would not be consistent with this scenario, and a change in its pitch angle, with $\psi\sim0$ after the Norma tangency would be needed to connect them (this was done in Fig.~\ref{fig:xyplane_last}, where the continuation of the Perseus arm is indicated with a black dashed line).

\paragraph{Outer arm}Regarding the structure of the Outer arm as it spirals inward into the first quadrant, the present distribution of CCs suggests that it might be instead connected to the Sct-Cen arm, possibly coinciding with the inner of the two gas component peaks present in the Centaurus tangency direction. This scenario would explain the double bump feature observed in this direction with different gas tracers \citep[see][and references therein]{hou+han2015}. The rest of the tangency points in the fourth quadrant are consistent with the spiral arm structure proposed here, as shown in Fig.1. In order to be able to extend the Outer arm from $\beta \sim -200\degr$ to $300\degr$ and at the same time be consistent with the near-side arm segments determined in R19, we needed to allow for a ‘‘kink’’ and change the pitch angle from $\sim10\degr$ to $7\degr$ (see Table~\ref{tab:Spiral_arm_param}).

The outward extension of the Outer arm proposed here is roughly consistent (in terms of its pitch-angle and spatial distribution) with the distant spiral arm in the fourth quadrant reported in \cite{mcclure-griffiths2004} traced in \ion{H}{i}. It would be desirable to study CCs in the region of the Outer arm traced by \cite{mcclure-griffiths2004} to put better constraints on the position of this arm on its way through the far side of the third and fourth Galactic quadrants. This would also allow to test if the extension proposed by us, with the Outer arm continuing to a second passage behind the Galactic center, at $R_{\mathrm{GC}}\gtrsim21\,\rm{kpc}$, is justified.

\paragraph{Scutum Centaurus arm}The Sct-Cen arm is one of the most clearly traced arms by our sample of spectroscopically classified young Cepheids. As shown in Fig.1, we support the association of the water maser source reported in \cite{sanna2017} with the Sct-Cen arm. \cite{sanna2017} suggested that this maser might link the accurately traced Sct-Cen arm segment in the near side of the first quadrant with a structure discovered by \cite{dame&thaddeus2011} in the far side of the first quadrant, detected in \ion{H}{i} studies \citep[see][and references therein]{koo2017} as well as traced with CO clouds \citep{dame&thaddeus2011,sun2015}. This link was based on the extrapolation of the Sct-Cen arm over a large section of the Galaxy. With the current data, we are able to add constraints on the passage of the Sct-Cen arm through the far side of the MW disk and provide further evidence that it runs over $\gtrsim 360\degr$ around the Galaxy. The pitch angle of $\sim13\degr$ adopted here is consistent with the values reported in the literature \citep{koo2017,reid2019} and the log-periodic spiral model passes near the Sct-Cen arm tangency at $\ell\approx306\degr$.

\paragraph{Sagittarius-Carina arm}For the Sgr-Car arm, we have 5 CCs tightly distributed around the proposed spiral arm model. Although these stars allow us to delineate a small segment of this spiral arm, when we add the CCs from the OGLE sample its structure becomes clearer. A pitch angle of $\sim12.7$\deg{} fits well the CCs positions, while still being consistent with a large portion of the arm segment traced with maser sources. This value is also compatible with the distribution of dense \ion{H}{i} gas associated to the Sgr-Car arm in the fourth quadrant of the face-on map recently produced by \cite{koo2017}. It may be needed to allow this arm for a kink in the near side of the first quadrant, somewhere in the range $15\degr\lesssim\beta\lesssim45\degr$, with a decrease in its pitch angle, in order for it to follow the positions of masers presented in R19 while at the same time being consistent with the Carina tangency direction and our CCs. Analogously to what we proposed for the Outer arm, the Sgr-Car arm may have a comparable (symmetric) origin, but as a branch of the Perseus arm, provided that we extrapolate the best fit model in the first quadrant from R19.

It is interesting to notice that, although departures from a ‘‘perfect’’ log-periodic spiral shape might exist, they do not appear to be large. A model with constant pitch angle seems sufficient to connect spiral arm tracers over most of the arm's extension, at least in the outer regions of the Galaxy.

There are not many CCs between the proposed arms on the far side of the disk, i.e., there seems to be a lack of CCs in the inter-arm regions (especially when the youngest CCs are considered). We would like to reinforce here that the proposed spiral arm model is not intended to represent a definitive picture of the far side spiral structure, but to show the potential that the use of CCs with distances based on near-IR photometry has on this field and how they can contribute to further constrain the far side disk spiral structure. The number of CCs used in the above analysis is still low and their distances are affected by our knowledge of the near-IR extinction law, a subject that is still debated in the literature \citep{nataf2016,majaess2016}. Enlarging the number of bona fide CCs in this region is desirable, in particular at $R_{\rm{GC}}\lesssim6-7\,\rm{kpc}$ and $R_{\rm{GC}}\gtrsim 12-14\,\rm{kpc}$.

\subsection{Tentative birthplaces of classical Cepheids assuming a rigid-body spiral pattern}\label{subsection:patternspeed}

In this section, we present the position of CCs, compared to spiral arms, under the assumption that CCs (like other stars) follow the disk rotation velocity, while spiral arms move as a rigid-body around the Galactic centre with a period of 250\,Myr. We therefore compute the current position of the birth-site of each CC by the difference between the disk rotation and the spiral pattern speed, integrated over the age of the CC, inferred from its period.

\begin{figure*}[ht]
\centering
  \includegraphics[width=17cm]{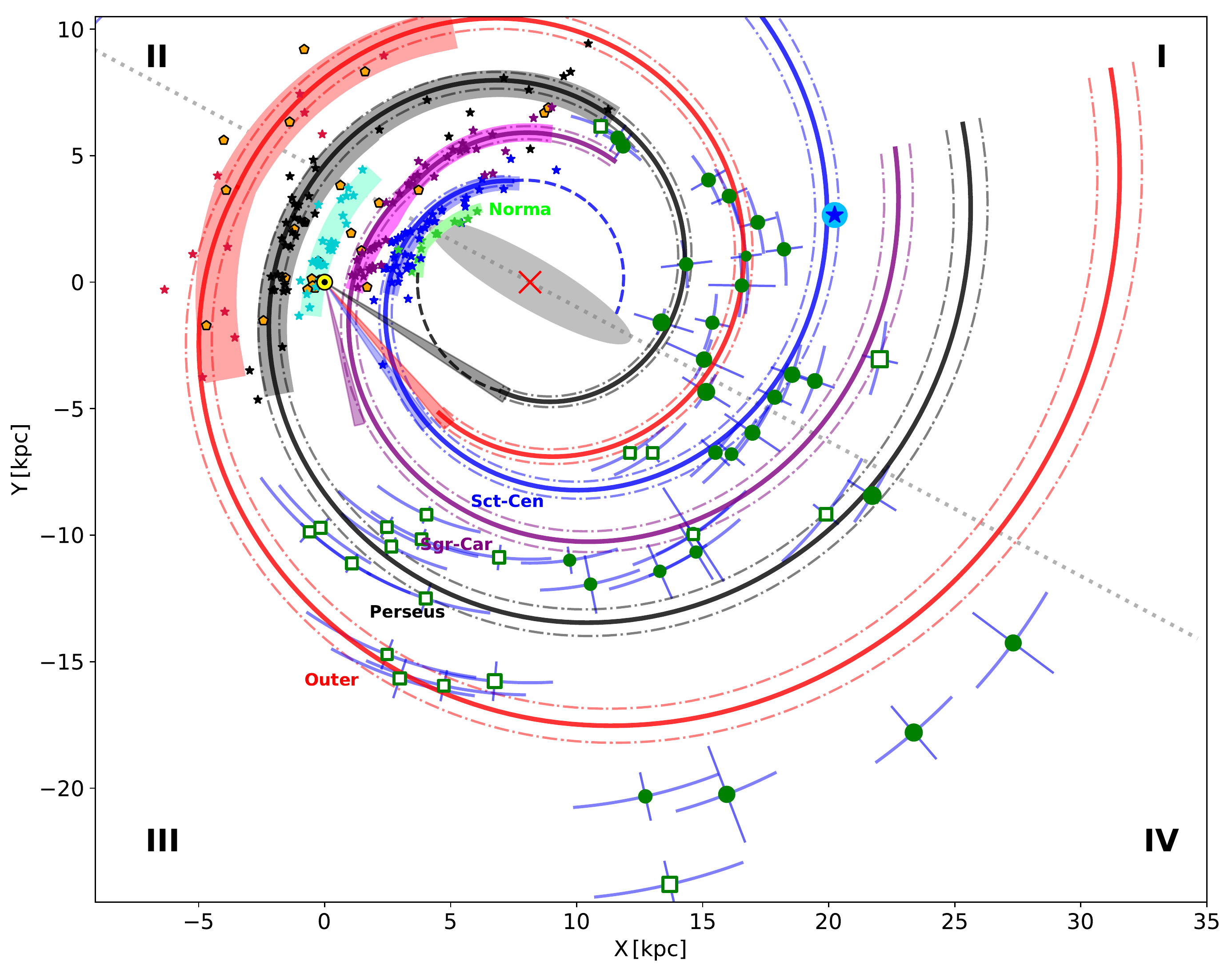}
       \caption{Distribution of our CCs birthplaces assuming they are born at the spiral arms and that the spiral pattern of our Galaxy has a fixed rotation period of 250\,Myr. We have rotated the observed positions of our CCs by the angular difference during their lifetimes between the assumed spiral pattern speed and \cite{mroz2019} rotation curve at their corresponding Galactocentric radii. The symbols are the same as in Fig.~\ref{fig:xyplane_last}. Individual error bars are indicated for each star (see the text for an explanation).}
       \label{fig:250Myr_pattern_speed}%
\end{figure*}

It is generally assumed in the literature that the spiral pattern moves at a fixed speed (rigid-body rotation). This assumption is related to the adoption of the quasi-stationary density wave theory \citep{lin&shu1964} as the mechanism that generates spiral structure. To test the effect of this hypothesis in the current position of our CCs with respect to their birthplaces, we adopt a pattern speed ($\Omega_p$) corresponding to a rotational period of $250\,\rm{Myr}$ \citep{vallee2017a} for the spiral arms ($\Omega_p \sim 24.5 \mathrm{\,km\,s^{-1}\,kpc^{-1}}$), as in \cite{skowron2019}. We thus calculated the current location of their birth-regions, by rotating the observed positions of our CCs. This was done taking into account the angular difference between both locations, originated from the differing angular speeds of the spiral pattern and the disk rotation \citep[for this purpose, we used the rotation curve from][]{mroz2019} at the corresponding Galactocentric radius during the CC lifetime. This transformation is shown in Fig.~\ref{fig:250Myr_pattern_speed}. We have estimated the uncertainty in the azimuthal direction based on the effect that an uncertainty in $\Omega_p$ of $ \pm 2\,\mathrm{km\,s^{-1}\,kpc^{-1}}$ would have in the computed positions; the radial component displayed here is the projected component of the heliocentric distance error bar shown in Fig.~\ref{fig:xyplane_errorbars}. While for the inner three arm segments the associated birthplaces preserve their consistency with our proposed spiral arm structure, the outermost arms clearly depart from the expected position of the birthplaces under this assumption. We tested other values reported in the literature for the spiral pattern speed \citep[see][]{gerhard2011,dias2019}, and obtained similar results, with the position of the inner arm segments proposed in subsection~\ref{subsection:xyplane} being roughly consistent with the computed birth-sites. As a cautionary note to this kind of modelling, we point out that the computed birthplace positions strongly depend on the assumed spiral pattern speed, and both the wide range of values reported in the literature and their underlying uncertainties prevent us from obtaining a new fit of the spiral arm structure under the rigid-body rotation scenario. It is also important to consider that there are a number of spiral arm formation mechanisms proposed in the literature and their predictions about the spiral pattern speed differ. N-body simulations often show a radially decaying pattern speed. As far as we know, neither the spiral pattern speed for the MW is well established over the full extent of the disk, nor the assumption that it rotates as a rigid-body has been definitely proven \citep[for a detailed discussion, see][]{dobbs&baba2014,pettitt2020}. Recently, based on the current positions and computed birthplaces of the most complete catalogue of Milky Way open clusters with Gaia EDR3 astrometry, \cite{castro-ginard2021} found that different spiral arms have distinct angular velocities, nearly co-rotating with the disk and disfavouring the density wave scenario. The understanding of the spiral structure of our Galaxy and the mechanisms involved in its formation are not yet settled.

In the previous subsection, we have presented the spatial distribution of our sample of spectroscopically confirmed CCs on the far disk. We notice that they follow arc-like distributions which we associate with the far side spiral structure. We found that there is a consistent picture between the near side spiral arm structure traced by the very young HMSFRs and our far side, long-period CCs for both the Perseus and Sct-Cen arms (Fig.~\ref{fig:xyplane_last}), suggesting that these stars might be reasonably good tracers of the spiral structure. When we add the CCs from the OGLE catalogue with VVV photometry, which are mainly shorter period, older CCs, we find that their in-plane distribution also agrees with this scenario, although showing a seemingly larger dispersion around the proposed arms. This preliminary analysis of the far side spiral structure using CCs will certainly improve when a larger sample of bona fide CCs becomes available in this region. The extension of the VVV survey \citep[VVVX,][]{minniti2018} will certainly play a major role in this sense, as will the Vera C. Rubin Observatory Legacy  Survey of Space and Time \citep[LSST,][]{ivezic2019_LSST}.

The combination of near-IR photometry (e.g., from the VVV survey) and near-IR spectroscopic follow-up has proven useful to obtain clean samples of these stars \citep{inno2019,minniti2020} in the far disk. This region remains out of reach for studies at shorter wavelengths and is also lacking other spiral arm tracers with accurate distance determinations. New VLBI arrays in the southern hemisphere will provide parallaxes for HMSFRs in the southern disk 
(R19; \citealt{vera2020}), allowing to test our results using CCs and to further constrain the MW spiral structure.

\subsection{Distribution perpendicular to the Galactic plane}\label{subsection:R-Z_plane}

\begin{figure*}
\sidecaption
  \includegraphics[width=12cm]{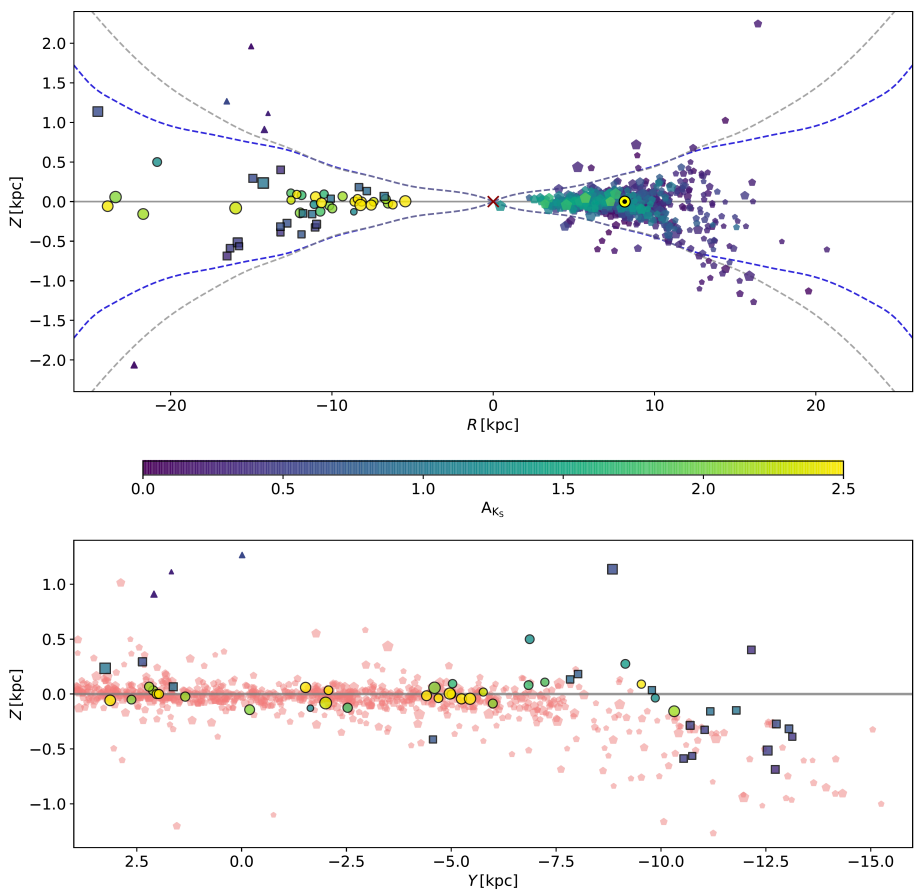}
     \caption{Top: Galactocentric radius ($R$) versus $Z$ distribution for CCs, color coded by their $K_{\rm{S}}$ extinctions, $A_{K_{\rm S}}$. At positive $R$ values, we plotted the CCs from \cite{skowron2019} that are on the near side of the Galactic disk, i.e., at $X<8.15\,\mathrm{kpc}$ (pentagons), while the CCs used in this work are shown at $R<0$ (circles and squares are used as in Fig. 1). The sizes of the symbols are inversely proportional to the CC ages. The five CCs from \cite{feast2014} are included here as a reference (triangles), as well as the flaring curve models from \cite{kalberla2007}, where the blue and grey dashed lines represent their S and N1 models, respectively. Bottom: Projection of our CCs in Galactic $Y$ versus $Z$ coordinates. The whole sample of stars from \cite{skowron2019b} with period > 5\,days is now plotted as background pink pentagons. The warped shape of the southern disk toward negative $Z$ values is clear in this plot.}
     \label{fig:z-R_plane}
\end{figure*}

In this section, we analyse the vertical distribution of our CCs. Fig.~\ref{fig:z-R_plane} presents their distances from the Galactic plane, $Z$, as a function of both their Galactocentric radii and $Y$ cartesian coordinates. As a comparison, we have included the CCs from \cite{skowron2019,skowron2019b} with period > 5\,days (to compare stars in a similar age-range).

It is clear that the OGLE stars added to our sample are further away from the plane (and preferentially at $Z<0\,\rm{kpc}$) than the spectroscopic sample. This is a result of the combined effect of at least two factors: Most of these OGLE CCs are located in the southern disk, at $l\lesssim340\degr$, where the Galactic warp lies below the plane\footnote{The line of nodes of the Galactic warp model by \cite{skowron2019} is shown in Fig.~\ref{fig:xyplane_last} (gray dotted line)} (bottom panel in Fig.~\ref{fig:z-R_plane}). The other aspect is the selection effect produced by extinction, preventing OGLE from finding far disk CCs close to the plane (note their $K_{\rm{S}}$ extinctions, color coded in Fig.~\ref{fig:z-R_plane}). This Figure also shows why near-IR observations are needed to study the far disk, where the VVV data can map high extinction regions (with $A_{K_{\rm S}} \geq 2.5$ mag).

Figure~\ref{fig:z-R_plane} shows that the full sample of CCs used in this work is tightly distributed around the Galactic mid-plane (note that the scale of the vertical axis is more expanded than the horizontal one). We find a planar distribution up to $R\sim24\,\rm{kpc}$, with a tighter $Z$ distribution compared to the flared structure mapped in \ion{H}{i} gas \citep{kalberla2007}, traced by Blue Straggler stars \citep{thomas2019}, or as suggested by the first CCs reported on the far side of the disk \citep{feast2014}, which do not appear to represent the distribution of the bulk of the CC variables present in the far disk.

Therefore, we find that there are numerous Cepheids in the far side of the Galaxy that are concentrated towards the mid-plane. From our observations, it does not appear that the far side of the MW is severely warped inside $R_\mathrm {GC} \sim 12\,\mathrm{kpc}$. On the other hand, if it was severely warped, we would not have found any CCs in the plane at those distances. However, our data are consistent with the warped structure observed by \cite{skowron2019,skowron2019b} at larger distances. The optical (OGLE) and near-IR (VVV) data complement each other nicely to give a consistent panorama of the distant side of the Galaxy.

\section{Discussion and conclusions}

There have been many previous attempts to map the spiral arm structure towards the inner region of the Galaxy, using different tracers, and at a variety of wavelengths. In particular, near-IR photometry has been widely used, allowing for example to study the tangency directions as traced by the distribution of the old Red Clump (RC) stars in the disk \citep{benjamin2008,hou+han2015}. Using VVV data, the far side of the disk has also been studied by means of RC giants through windows of low extinction across the Galactic plane \citep{minniti2018,gonzalez2018,saito2020}. Reaching distances up to $\sim 14\,\rm{kpc}$, these studies have found evidence of the spiral arm structure as traced by these old stars.

In this work we employ CCs, which are bright, young, and follow accurate PL relations, making them potentially good tracers of the spiral arm structure. We combined IR and optically classified samples of bona fide CCs and homogeneously determined their distances based on near-IR photometry from the VVV survey. Our stars are located in the Galactic plane at longitudes that range from $\sim+10\degr$ to $-65\degr$, allowing us to map a region of the disk that is poorly studied and where kinematic distances, widely used for other tracers observable in the radio regime for example, are degenerate.

In previous works, the arm segments known on our side of the Galaxy were connected with each other (e.g., Norma-Outer arm) or to the few known characteristic on the far side \citep[like the maser by][]{sanna2017} by extrapolating over large distances. CCs potentially will enable us to test the extrapolation of the MW spiral arms in the poorly explored disk far side. In addition, with a larger statistical sample we might be able to put robust constrains on the spiral pattern rotation.

With minor modifications to the near-side spiral arm structure traced by maser parallaxes of HMSFRs, the current position of the present CC sample can be fitted into a coherent picture of the MW spiral arms (Fig.~\ref{fig:xyplane_last}). On the contrary, a tentative correction for the different rotation of the stars -- integrated over the age of each CC -- and the spiral pattern speed provides a less clear picture (Fig.~\ref{fig:250Myr_pattern_speed}). Although the number of CCs is still small, this might be taken as evidence favouring the hypothesis that these stars have not significantly drifted away from their birthplaces, possibly moving together with the arms and against the idea of spiral arms rotating as a rigid-body in the Milky Way. Therefore in what follows we will discuss the position of the spiral arms as constrained by the current positions of CCs, shown in Fig.~\ref{fig:xyplane_last}. Note that in M31 \cite{kodric2018} showed a similar result, namely, that the current location of CCs coincides well with the spiral- and ring-like structures traced by dust (see their Fig.~17).

By means of the youngest CCs, that are still close to their birth locations, we propose a tentative location of the spiral arms in the far disk that is compatible with local tracers. Our data suggests that the Perseus arm is connected with the so-called Norma tangency point. We also add support to the continuation of the Sct-Cen arm beyond the Galactic center as suggested by \cite{dame&thaddeus2011}. The position of the maser reported in \cite{sanna2017} is fully consistent with the distribution of CCs, which allow us to extend the portion of this arm traced by young stellar objects. The present data do not favour the Outer and Norma arms being connected. We speculate that the Outer arm might be connected with the second peak known to exist in the Centaurus arm tangency direction. In this scenario, the Outer arm would be a ‘‘branch’’ of the Sct-Cen arm. If extrapolated, an analogous origin with respect to the Perseus arm appears to be plausible for the Sgr-Car arm as traced by masers. This would give further support to the symmetry between the Sct-Cen and Perseus arms \citep{dame&thaddeus2011}.

\begin{figure}[ht]
\includegraphics[width=\columnwidth]{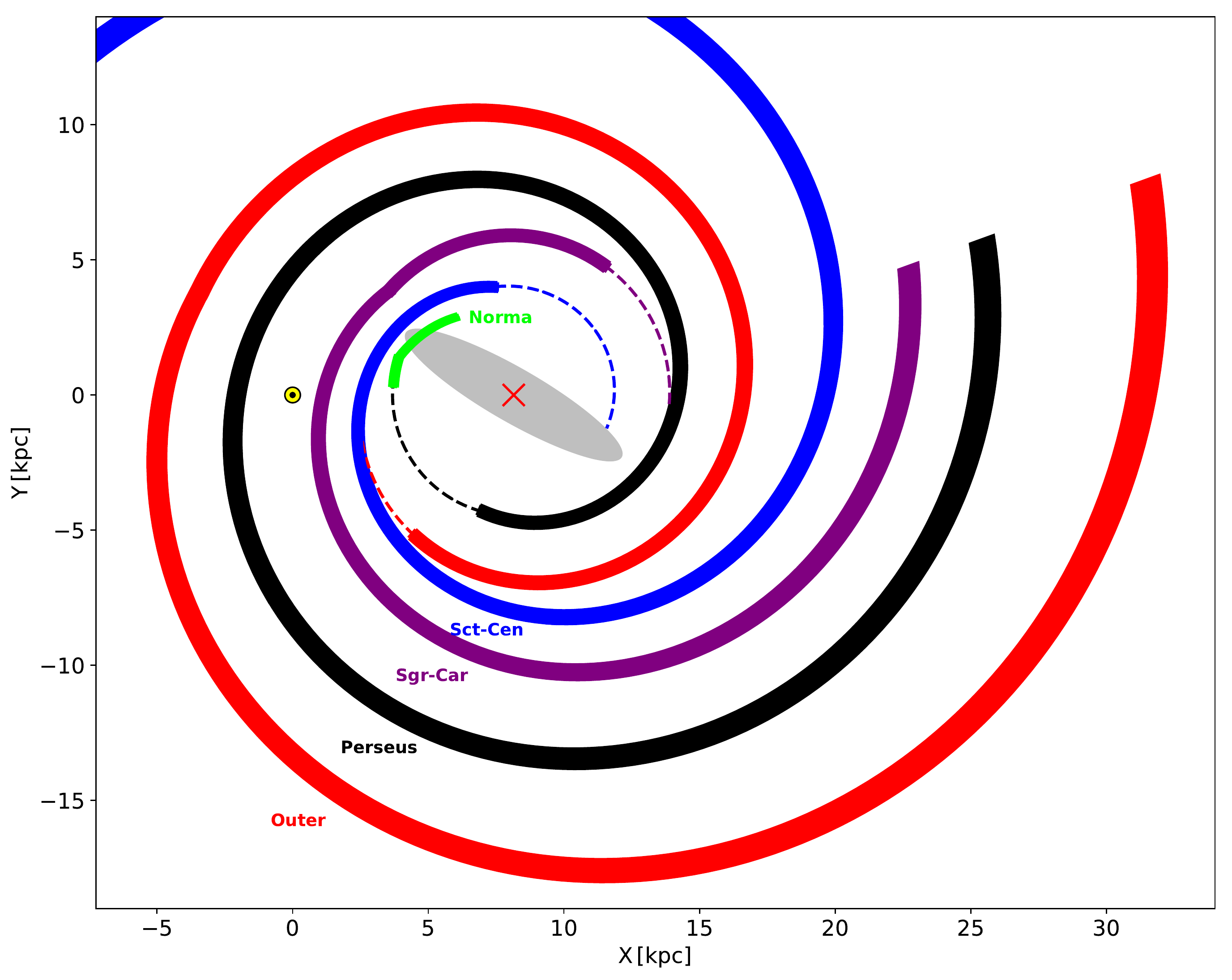}
\caption{Schematic representation of the spiral arm model of the MW as proposed in this work. The Sun is marked with a yellow dot and the Galactic center with a red cross. The dashed lines are arm extrapolations; for the Perseus arm, we changed the pitch angle of this inner segment to $\psi=0\degr$ in order to connect it with the Norma arm, and for the Sct-Cen arm the pitch angle was set to $\psi=3\degr$ to connect it with the other extreme of the bar.}
\label{fig:proposed_spiral_structure}
\end{figure}

If this is the case, then the MW would have 2 major arms, branching out into four arms in the outer regions as traced by young stars (see Fig.~\ref{fig:proposed_spiral_structure} for a schematic representation of the proposed spiral structure). This could resolve the seemingly inconsistent observational evidence in the near-IR pointing to a two-arm spiral pattern as traced by old stars \citep[e.g.][]{benjamin2008}, with the results obtained here and from VLBI maser observations, as has been already suggested for example by \cite{reid2009}. The apparent lack of CCs in the inter-arm regions is worth further attention. If this persists as the sample is expanded, it would provide valuable constraints on the possible arrangement of the arms. Our CC sample indicates that the spiral arms of the MW revolve more than $\sim 360\degr$ around the Galactic centre and their distribution is confined to the plane of the Galaxy.

Our somewhat simple model (which only requires for a kink in the Outer arm) is qualitatively consistent with the present data and with HMSFRs positions. The CCs presented here provide a prospect of the constraints that the continued expansion of the sample size (and mitigated uncertainties) could yield on the spiral structure of the far side of the disk, though they are still a relatively small statistical sample to allow a robust tracing of the spiral arms. We believe that using CCs we will be able to extend the spiral arm segments in the near side of the disk to the far side on the first and fourth Galactic quadrants. The study of the far Galactic spiral structure would be greatly benefited by the assemblage of a larger sample of bona fide CCs in this area. They are a promising tool to provide us with a clearer and global picture of the spiral arms, complementing the current and coming VLBI and {\em Gaia} studies of the youngest stellar populations in the disk, that combined will allow to improve our knowledge about the spiral structure of our home Galaxy.

\begin{acknowledgements}
We gratefully acknowledge data from the ESO Public Survey program ID 179.B-2002 taken with the VISTA telescope, and products from the Cambridge Astronomical Survey Unit (CASU) and from the VISTA Science Archive (VSA). Support for this work was provided by the BASAL Center for Astrophysics and Associated Technologies (CATA) through grant AFB170002, and the Ministry for the Economy, Development and Tourism, Programa Iniciativa Cient{\'i}fica Milenio grant IC120009, awarded to the Millennium Institute of Astrophysics (MAS). This work is part of the Ph.D. thesis of J.H.M., funded by grant CONICYT-PCHA Doctorado Nacional 2015-21151640. A significant part of this work was performed when J.H.M. was affiliated to ESO. J.H.M. thanks ESO for the studentship at ESO Santiago. We acknowledge additional support by Proyecto Fondecyt Regular \#1191505 (M.Z.), \#1171273 (M.C.), \#1170121 (D.M.). A.R.A acknowledges partial support from FONDECYT through grant \#3180203. This research has made use of NASA’s Astrophysics Data System Bibliographic Services. This work made use of PYTHON routines in the {\tt astropy} package \citep{astropy13}.
\end{acknowledgements}

\bibliographystyle{aa}
\bibliography{biblio}

\begin{thebibliography}{57}
\expandafter\ifx\csname natexlab\endcsname\relax\def\natexlab#1{#1}\fi

\bibitem[{{Astropy Collaboration} {et~al.}(2013){Astropy Collaboration},
  {Robitaille}, {Tollerud}, {Greenfield}, {Droettboom}, {Bray}, {Aldcroft},
  {Davis}, {Ginsburg}, {Price-Whelan}, {Kerzendorf}, {Conley}, {Crighton},
  {Barbary}, {Muna}, {Ferguson}, {Grollier}, {Parikh}, {Nair}, {Unther},
  {Deil}, {Woillez}, {Conseil}, {Kramer}, {Turner}, {Singer}, {Fox}, {Weaver},
  {Zabalza}, {Edwards}, {Azalee Bostroem}, {Burke}, {Casey}, {Crawford},
  {Dencheva}, {Ely}, {Jenness}, {Labrie}, {Lim}, {Pierfederici}, {Pontzen},
  {Ptak}, {Refsdal}, {Servillat}, \& {Streicher}}]{astropy13}
{Astropy Collaboration}, {Robitaille}, T.~P., {Tollerud}, E.~J., {et~al.} 2013,
  \aap, 558, A33

\bibitem[{{Benjamin}(2008)}]{benjamin2008}
{Benjamin}, R.~A. 2008, in Astronomical Society of the Pacific Conference
  Series, Vol. 387, Massive Star Formation: Observations Confront Theory, ed.
  H.~{Beuther}, H.~{Linz}, \& T.~{Henning}, 375

\bibitem[{{Bland-Hawthorn} \& {Gerhard}(2016)}]{blandhawthorn16}
{Bland-Hawthorn}, J. \& {Gerhard}, O. 2016, \araa, 54, 529

\bibitem[{{Castro-Ginard} {et~al.}(2021){Castro-Ginard}, {McMillan}, {Luri},
  {Jordi}, {Romero-G{\'o}mez}, {Cantat-Gaudin}, {Casamiquela}, {Tarricq},
  {Soubiran}, \& {Anders}}]{castro-ginard2021}
{Castro-Ginard}, A., {McMillan}, P.~J., {Luri}, X., {et~al.} 2021, arXiv
  e-prints, arXiv:2105.04590

\bibitem[{{Catelan} \& {Smith}(2015)}]{catelan2015}
{Catelan}, M. \& {Smith}, H.~A. 2015, {Pulsating Stars} (Wiley-VCH, Weinheim)

\bibitem[{{Chen} {et~al.}(2019){Chen}, {Huang}, {Hou}, {Tian}, {Li}, {Yuan},
  {Wang}, {Wang}, {Tian}, \& {Liu}}]{chen2019}
{Chen}, B.~Q., {Huang}, Y., {Hou}, L.~G., {et~al.} 2019, \mnras, 487, 1400

\bibitem[{{Dame} \& {Thaddeus}(2011)}]{dame&thaddeus2011}
{Dame}, T.~M. \& {Thaddeus}, P. 2011, \apjl, 734, L24

\bibitem[{{De Somma} {et~al.}(2020){De Somma}, {Marconi}, {Cassisi}, {Ripepi},
  {Leccia}, {Molinaro}, \& {Musella}}]{desomma2020}
{De Somma}, G., {Marconi}, M., {Cassisi}, S., {et~al.} 2020, \mnras, 496, 5039

\bibitem[{{D{\'e}k{\'a}ny} {et~al.}(2019){D{\'e}k{\'a}ny}, {Hajdu}, {Grebel},
  \& {Catelan}}]{dekany2019}
{D{\'e}k{\'a}ny}, I., {Hajdu}, G., {Grebel}, E.~K., \& {Catelan}, M. 2019,
  \apj, 883, 58

\bibitem[{{Dias} {et~al.}(2019){Dias}, {Monteiro}, {L{\'e}pine}, \&
  {Barros}}]{dias2019}
{Dias}, W.~S., {Monteiro}, H., {L{\'e}pine}, J.~R.~D., \& {Barros}, D.~A. 2019,
  \mnras, 486, 5726

\bibitem[{{Dobbs} \& {Baba}(2014)}]{dobbs&baba2014}
{Dobbs}, C. \& {Baba}, J. 2014, \pasa, 31, e035

\bibitem[{{Feast} {et~al.}(2014){Feast}, {Menzies}, {Matsunaga}, \&
  {Whitelock}}]{feast2014}
{Feast}, M.~W., {Menzies}, J.~W., {Matsunaga}, N., \& {Whitelock}, P.~A. 2014,
  \nat, 509, 342

\bibitem[{{Genovali} {et~al.}(2014){Genovali}, {Lemasle}, {Bono}, {Romaniello},
  {Fabrizio}, {Ferraro}, {Iannicola}, {Laney}, {Nonino}, {Bergemann},
  {Buonanno}, {Fran{\c{c}}ois}, {Inno}, {Kudritzki}, {Matsunaga}, {Pedicelli},
  {Primas}, \& {Th{\'e}venin}}]{genovali2014}
{Genovali}, K., {Lemasle}, B., {Bono}, G., {et~al.} 2014, \aap, 566, A37

\bibitem[{{Gerhard}(2011)}]{gerhard2011}
{Gerhard}, O. 2011, Memorie della Societa Astronomica Italiana Supplementi, 18,
  185

\bibitem[{{Gonzalez} {et~al.}(2018){Gonzalez}, {Minniti}, {Valenti},
  {Alonso-Garc{\'\i}a}, {Debattista}, {Zoccali}, {Rejkuba}, {Dias}, {Surot},
  {Hempel}, \& {Saito}}]{gonzalez2018}
{Gonzalez}, O.~A., {Minniti}, D., {Valenti}, E., {et~al.} 2018, \mnras, 481,
  L130

\bibitem[{{Gravity Collaboration} {et~al.}(2019){Gravity Collaboration},
  {Abuter}, {Amorim}, {Baub{\"o}ck}, {Berger}, {Bonnet}, {Brand ner},
  {Cl{\'e}net}, {Coud{\'e} Du Foresto}, {de Zeeuw}, {Dexter}, {Duvert},
  {Eckart}, {Eisenhauer}, {F{\"o}rster Schreiber}, {Garcia}, {Gao}, {Gendron},
  {Genzel}, {Gerhard}, {Gillessen}, {Habibi}, {Haubois}, {Henning}, {Hippler},
  {Horrobin}, {Jim{\'e}nez-Rosales}, {Jocou}, {Kervella}, {Lacour},
  {Lapeyr{\`e}re}, {Le Bouquin}, {L{\'e}na}, {Ott}, {Paumard}, {Perraut},
  {Perrin}, {Pfuhl}, {Rabien}, {Rodriguez Coira}, {Rousset}, {Scheithauer},
  {Sternberg}, {Straub}, {Straubmeier}, {Sturm}, {Tacconi}, {Vincent}, {von
  Fellenberg}, {Waisberg}, {Widmann}, {Wieprecht}, {Wiezorrek}, {Woillez}, \&
  {Yazici}}]{gravity2019}
{Gravity Collaboration}, {Abuter}, R., {Amorim}, A., {et~al.} 2019, \aap, 625,
  L10

\bibitem[{{Hajdu} {et~al.}(2019){Hajdu}, {D{\'e}k{\'a}ny}, {Catelan}, \&
  {Grebel}}]{hajdu2019}
{Hajdu}, G., {D{\'e}k{\'a}ny}, I., {Catelan}, M., \& {Grebel}, E.~K. 2019,
  arXiv e-prints, arXiv:1908.06160

\bibitem[{{Hou} \& {Han}(2015)}]{hou+han2015}
{Hou}, L.~G. \& {Han}, J.~L. 2015, \mnras, 454, 626

\bibitem[{{Inno} {et~al.}(2016){Inno}, {Bono}, {Matsunaga}, {Fiorentino},
  {Marconi}, {Lemasle}, {da Silva}, {Soszy{\'n}ski}, {Udalski}, {Romaniello},
  \& {Rix}}]{inno2016}
{Inno}, L., {Bono}, G., {Matsunaga}, N., {et~al.} 2016, \apj, 832, 176

\bibitem[{{Inno} {et~al.}(2019){Inno}, {Urbaneja}, {Matsunaga}, {Bono},
  {Nonino}, {Debattista}, {Sormani}, {Bergemann}, {da Silva}, {Lemasle},
  {Romaniello}, \& {Rix}}]{inno2019}
{Inno}, L., {Urbaneja}, M.~A., {Matsunaga}, N., {et~al.} 2019, \mnras, 482, 83

\bibitem[{{Ivezi{\'c}} {et~al.}(2019){Ivezi{\'c}}, {Kahn}, {Tyson}, {Abel},
  {Acosta}, {Allsman}, {Alonso}, {AlSayyad}, {Anderson}, {Andrew}, {Angel},
  {Angeli}, {Ansari}, {Antilogus}, {Araujo}, {Armstrong}, {Arndt}, {Astier},
  {Aubourg}, {Auza}, {Axelrod}, {Bard}, {Barr}, {Barrau}, {Bartlett}, {Bauer},
  {Bauman}, {Baumont}, {Bechtol}, {Bechtol}, {Becker}, {Becla}, {Beldica},
  {Bellavia}, {Bianco}, {Biswas}, {Blanc}, {Blazek}, {Blandford}, {Bloom},
  {Bogart}, {Bond}, {Booth}, {Borgland}, {Borne}, {Bosch}, {Boutigny},
  {Brackett}, {Bradshaw}, {Brandt}, {Brown}, {Bullock}, {Burchat}, {Burke},
  {Cagnoli}, {Calabrese}, {Callahan}, {Callen}, {Carlin}, {Carlson},
  {Chandrasekharan}, {Charles-Emerson}, {Chesley}, {Cheu}, {Chiang}, {Chiang},
  {Chirino}, {Chow}, {Ciardi}, {Claver}, {Cohen-Tanugi}, {Cockrum}, {Coles},
  {Connolly}, {Cook}, {Cooray}, {Covey}, {Cribbs}, {Cui}, {Cutri}, {Daly},
  {Daniel}, {Daruich}, {Daubard}, {Daues}, {Dawson}, {Delgado}, {Dellapenna},
  {de Peyster}, {de Val-Borro}, {Digel}, {Doherty}, {Dubois},
  {Dubois-Felsmann}, {Durech}, {Economou}, {Eifler}, {Eracleous}, {Emmons},
  {Fausti Neto}, {Ferguson}, {Figueroa}, {Fisher-Levine}, {Focke}, {Foss},
  {Frank}, {Freemon}, {Gangler}, {Gawiser}, {Geary}, {Gee}, {Geha}, {Gessner},
  {Gibson}, {Gilmore}, {Glanzman}, {Glick}, {Goldina}, {Goldstein}, {Goodenow},
  {Graham}, {Gressler}, {Gris}, {Guy}, {Guyonnet}, {Haller}, {Harris},
  {Hascall}, {Haupt}, {Hernandez}, {Herrmann}, {Hileman}, {Hoblitt}, {Hodgson},
  {Hogan}, {Howard}, {Huang}, {Huffer}, {Ingraham}, {Innes}, {Jacoby}, {Jain},
  {Jammes}, {Jee}, {Jenness}, {Jernigan}, {Jevremovi{\'c}}, {Johns}, {Johnson},
  {Johnson}, {Jones}, {Juramy-Gilles}, {Juri{\'c}}, {Kalirai}, {Kallivayalil},
  {Kalmbach}, {Kantor}, {Karst}, {Kasliwal}, {Kelly}, {Kessler}, {Kinnison},
  {Kirkby}, {Knox}, {Kotov}, {Krabbendam}, {Krughoff}, {Kub{\'a}nek},
  {Kuczewski}, {Kulkarni}, {Ku}, {Kurita}, {Lage}, {Lambert}, {Lange},
  {Langton}, {Le Guillou}, {Levine}, {Liang}, {Lim}, {Lintott}, {Long},
  {Lopez}, {Lotz}, {Lupton}, {Lust}, {MacArthur}, {Mahabal}, {Mandelbaum},
  {Markiewicz}, {Marsh}, {Marshall}, {Marshall}, {May}, {McKercher}, {McQueen},
  {Meyers}, {Migliore}, {Miller}, {Mills}, {Miraval}, {Moeyens}, {Moolekamp},
  {Monet}, {Moniez}, {Monkewitz}, {Montgomery}, {Morrison}, {Mueller},
  {Muller}, {Mu{\~n}oz Arancibia}, {Neill}, {Newbry}, {Nief}, {Nomerotski},
  {Nordby}, {O'Connor}, {Oliver}, {Olivier}, {Olsen}, {O'Mullane}, {Ortiz},
  {Osier}, {Owen}, {Pain}, {Palecek}, {Parejko}, {Parsons}, {Pease},
  {Peterson}, {Peterson}, {Petravick}, {Libby Petrick}, {Petry},
  {Pierfederici}, {Pietrowicz}, {Pike}, {Pinto}, {Plante}, {Plate}, {Plutchak},
  {Price}, {Prouza}, {Radeka}, {Rajagopal}, {Rasmussen}, {Regnault}, {Reil},
  {Reiss}, {Reuter}, {Ridgway}, {Riot}, {Ritz}, {Robinson}, {Roby}, {Roodman},
  {Rosing}, {Roucelle}, {Rumore}, {Russo}, {Saha}, {Sassolas}, {Schalk},
  {Schellart}, {Schindler}, {Schmidt}, {Schneider}, {Schneider}, {Schoening},
  {Schumacher}, {Schwamb}, {Sebag}, {Selvy}, {Sembroski}, {Seppala}, {Serio},
  {Serrano}, {Shaw}, {Shipsey}, {Sick}, {Silvestri}, {Slater}, {Smith},
  {Smith}, {Sobhani}, {Soldahl}, {Storrie-Lombardi}, {Stover}, {Strauss},
  {Street}, {Stubbs}, {Sullivan}, {Sweeney}, {Swinbank}, {Szalay}, {Takacs},
  {Tether}, {Thaler}, {Thayer}, {Thomas}, {Thornton}, {Thukral}, {Tice},
  {Trilling}, {Turri}, {Van Berg}, {Vanden Berk}, {Vetter}, {Virieux},
  {Vucina}, {Wahl}, {Walkowicz}, {Walsh}, {Walter}, {Wang}, {Wang}, {Warner},
  {Wiecha}, {Willman}, {Winters}, {Wittman}, {Wolff}, {Wood-Vasey}, {Wu},
  {Xin}, {Yoachim}, \& {Zhan}}]{ivezic2019_LSST}
{Ivezi{\'c}}, {\v{Z}}., {Kahn}, S.~M., {Tyson}, J.~A., {et~al.} 2019, \apj,
  873, 111

\bibitem[{{Kalberla} {et~al.}(2007){Kalberla}, {Dedes}, {Kerp}, \&
  {Haud}}]{kalberla2007}
{Kalberla}, P.~M.~W., {Dedes}, L., {Kerp}, J., \& {Haud}, U. 2007, \aap, 469,
  511

\bibitem[{{Kodric} {et~al.}(2018){Kodric}, {Riffeser}, {Hopp}, {Goessl},
  {Seitz}, {Bender}, {Koppenhoefer}, {Obermeier}, {Snigula}, {Lee}, {Burgett},
  {Draper}, {Hodapp}, {Kaiser}, {Kudritzki}, {Metcalfe}, {Tonry}, \&
  {Wainscoat}}]{kodric2018}
{Kodric}, M., {Riffeser}, A., {Hopp}, U., {et~al.} 2018, \aj, 156, 130

\bibitem[{{Koo} {et~al.}(2017){Koo}, {Park}, {Kim}, {Lee}, {Balser}, \&
  {Wenger}}]{koo2017}
{Koo}, B.-C., {Park}, G., {Kim}, W.-T., {et~al.} 2017, \pasp, 129, 094102

\bibitem[{{Lemasle} {et~al.}(2018){Lemasle}, {Hajdu}, {Kovtyukh}, {Inno},
  {Grebel}, {Catelan}, {Bono}, {Fran{\c{c}}ois}, {Kniazev}, \& {da
  Silva}}]{lemasle2018}
{Lemasle}, B., {Hajdu}, G., {Kovtyukh}, V., {et~al.} 2018, \aap, 618, A160

\bibitem[{{Lin} \& {Shu}(1964)}]{lin&shu1964}
{Lin}, C.~C. \& {Shu}, F.~H. 1964, \apj, 140, 646

\bibitem[{{Luck}(2018)}]{luck2018}
{Luck}, R.~E. 2018, \aj, 156, 171

\bibitem[{{Macri} {et~al.}(2015){Macri}, {Ngeow}, {Kanbur}, {Mahzooni}, \&
  {Smitka}}]{macri2015}
{Macri}, L.~M., {Ngeow}, C.-C., {Kanbur}, S.~M., {Mahzooni}, S., \& {Smitka},
  M.~T. 2015, \aj, 149, 117

\bibitem[{{Majaess} {et~al.}(2016){Majaess}, {Turner}, {D{\'e}k{\'a}ny},
  {Minniti}, \& {Gieren}}]{majaess2016}
{Majaess}, D., {Turner}, D., {D{\'e}k{\'a}ny}, I., {Minniti}, D., \& {Gieren},
  W. 2016, \aap, 593, A124

\bibitem[{{McClure-Griffiths} {et~al.}(2004){McClure-Griffiths}, {Dickey},
  {Gaensler}, \& {Green}}]{mcclure-griffiths2004}
{McClure-Griffiths}, N.~M., {Dickey}, J.~M., {Gaensler}, B.~M., \& {Green},
  A.~J. 2004, \apjl, 607, L127

\bibitem[{{Minniti} {et~al.}(2010){Minniti}, {Lucas}, {Emerson}, {Saito},
  {Hempel}, {Pietrukowicz}, {Ahumada}, {Alonso}, {Alonso-Garcia}, \&
  {Arias}}]{minniti2010}
{Minniti}, D., {Lucas}, P.~W., {Emerson}, J.~P., {et~al.} 2010, New Astronomy,
  15, 433

\bibitem[{{Minniti} {et~al.}(2018){Minniti}, {Saito}, {Gonzalez},
  {Alonso-Garc{\'\i}a}, {Rejkuba}, {Barb{\'a}}, {Irwin}, {Kammers}, {Lucas},
  {Majaess}, \& {Valenti}}]{minniti2018}
{Minniti}, D., {Saito}, R.~K., {Gonzalez}, O.~A., {et~al.} 2018, \aap, 616, A26

\bibitem[{{Minniti} {et~al.}(2020){Minniti}, {Sbordone}, {Rojas-Arriagada},
  {Zoccali}, {Contreras Ramos}, {Minniti}, {Marconi}, {Braga}, {Catelan},
  {Duffau}, {Gieren}, \& {Valcarce}}]{minniti2020}
{Minniti}, J.~H., {Sbordone}, L., {Rojas-Arriagada}, A., {et~al.} 2020, \aap,
  640, A92

\bibitem[{{Mr{\'o}z} {et~al.}(2019){Mr{\'o}z}, {Udalski}, {Skowron}, {Skowron},
  {Soszy{\'n}ski}, {Pietrukowicz}, {Szyma{\'n}ski}, {Poleski}, {Koz{\l}owski},
  \& {Ulaczyk}}]{mroz2019}
{Mr{\'o}z}, P., {Udalski}, A., {Skowron}, D.~M., {et~al.} 2019, \apj, 870, L10

\bibitem[{{Nataf} {et~al.}(2016){Nataf}, {Gonzalez}, {Casagrande}, {Zasowski},
  {Wegg}, {Wolf}, {Kunder}, {Alonso-Garcia}, {Minniti}, {Rejkuba}, {Saito},
  {Valenti}, {Zoccali}, {Poleski}, {Pietrzy{\'n}ski}, {Skowron},
  {Soszy{\'n}ski}, {Szyma{\'n}ski}, {Udalski}, {Ulaczyk}, \&
  {Wyrzykowski}}]{nataf2016}
{Nataf}, D.~M., {Gonzalez}, O.~A., {Casagrande}, L., {et~al.} 2016, \mnras,
  456, 2692

\bibitem[{{Pettitt} {et~al.}(2020){Pettitt}, {Ragan}, \& {Smith}}]{pettitt2020}
{Pettitt}, A.~R., {Ragan}, S.~E., \& {Smith}, M.~C. 2020, \mnras, 491, 2162

\bibitem[{{Reid} {et~al.}(2019){Reid}, {Menten}, {Brunthaler}, {Zheng}, {Dame},
  {Xu}, {Li}, {Sakai}, {Wu}, {Immer}, {Zhang}, {Sanna}, {Moscadelli}, {Rygl},
  {Bartkiewicz}, {Hu}, {Quiroga-Nu{\~n}ez}, \& {van Langevelde}}]{reid2019}
{Reid}, M.~J., {Menten}, K.~M., {Brunthaler}, A., {et~al.} 2019, \apj, 885, 131

\bibitem[{{Reid} {et~al.}(2014){Reid}, {Menten}, {Brunthaler}, {Zheng}, {Dame},
  {Xu}, {Wu}, {Zhang}, {Sanna}, {Sato}, {Hachisuka}, {Choi}, {Immer},
  {Moscadelli}, {Rygl}, \& {Bartkiewicz}}]{reid2014}
{Reid}, M.~J., {Menten}, K.~M., {Brunthaler}, A., {et~al.} 2014, \apj, 783, 130

\bibitem[{{Reid} {et~al.}(2009){Reid}, {Menten}, {Zheng}, {Brunthaler},
  {Moscadelli}, {Xu}, {Zhang}, {Sato}, {Honma}, {Hirota}, {Hachisuka}, {Choi},
  {Moellenbrock}, \& {Bartkiewicz}}]{reid2009}
{Reid}, M.~J., {Menten}, K.~M., {Zheng}, X.~W., {et~al.} 2009, \apj, 700, 137

\bibitem[{{Ripepi} {et~al.}(2020){Ripepi}, {Catanzaro}, {Molinaro}, {Marconi},
  {Clementini}, {Cusano}, {De Somma}, {Leccia}, {Musella}, \&
  {Testa}}]{ripepi2020}
{Ripepi}, V., {Catanzaro}, G., {Molinaro}, R., {et~al.} 2020, arXiv e-prints,
  arXiv:2008.04608

\bibitem[{{Ripepi} {et~al.}(2017){Ripepi}, {Cioni}, {Moretti}, {Marconi},
  {Bekki}, {Clementini}, {de Grijs}, {Emerson}, {Groenewegen}, {Ivanov},
  {Molinaro}, {Muraveva}, {Oliveira}, {Piatti}, {Subramanian}, \& {van
  Loon}}]{ripepi2017}
{Ripepi}, V., {Cioni}, M.-R.~L., {Moretti}, M.~I., {et~al.} 2017, \mnras, 472,
  808

\bibitem[{{Saito} {et~al.}(2020){Saito}, {Minniti}, {Benjamin}, {Navarro},
  {Alonso-Garc{\'\i}a}, {Gonzalez}, {Kammers}, \& {Surot}}]{saito2020}
{Saito}, R.~K., {Minniti}, D., {Benjamin}, R.~A., {et~al.} 2020, \mnras, 494,
  L32

\bibitem[{{Sanna} {et~al.}(2017){Sanna}, {Reid}, {Dame}, {Menten}, \&
  {Brunthaler}}]{sanna2017}
{Sanna}, A., {Reid}, M.~J., {Dame}, T.~M., {Menten}, K.~M., \& {Brunthaler}, A.
  2017, Science, 358, 227

\bibitem[{Skowron {et~al.}(2019{\natexlab{a}})Skowron, Skowron, Mr{\'o}z,
  Udalski, Pietrukowicz, Soszy{\'n}ski, Szyma{\'n}ski, Poleski, Koz{\l}owski,
  Ulaczyk, Rybicki, \& Iwanek}]{skowron2019}
Skowron, D.~M., Skowron, J., Mr{\'o}z, P., {et~al.} 2019{\natexlab{a}},
  Science, 365, 478

\bibitem[{Skowron {et~al.}(2019{\natexlab{b}})Skowron, {Skowron}, {Mr{\'o}z},
  {Udalski}, {Pietrukowicz}, {Soszy{\'n}ski}, {Szyma{\'n}ski}, {Poleski},
  {Koz{\l}owski}, {Ulaczyk}, {Rybicki}, {Iwanek}, {. Wrona}, \&
  {Gromadzki}}]{skowron2019b}
Skowron, D.~M., {Skowron}, J., {Mr{\'o}z}, P., {et~al.} 2019{\natexlab{b}},
  \actaa, 69, 305

\bibitem[{{Soszy{\'n}ski} {et~al.}(2020){Soszy{\'n}ski}, {Udalski},
  {Szyma{\'n}ski}, {Pietrukowicz}, {Skowron}, {Skowron}, {Poleski},
  {Koz{\l}owski}, {Mr{\'o}z}, {Ulaczyk}, {Rybicki}, {Iwanek}, {Wrona}, \&
  {Gromadzki}}]{soszynski2020}
{Soszy{\'n}ski}, I., {Udalski}, A., {Szyma{\'n}ski}, M.~K., {et~al.} 2020,
  \actaa, 70, 101

\bibitem[{{Sun} {et~al.}(2015){Sun}, {Xu}, {Yang}, {Li}, {Du}, {Zhang}, \&
  {Zhou}}]{sun2015}
{Sun}, Y., {Xu}, Y., {Yang}, J., {et~al.} 2015, \apjl, 798, L27

\bibitem[{{Thomas} {et~al.}(2019){Thomas}, {Laporte}, {McConnachie}, {Famaey},
  {Ibata}, {Martin}, {Starkenburg}, {Carlberg}, {Malhan}, \&
  {Venn}}]{thomas2019}
{Thomas}, G.~F., {Laporte}, C. F.~P., {McConnachie}, A.~W., {et~al.} 2019,
  \mnras, 483, 3119

\bibitem[{{Udalski} {et~al.}(2018){Udalski}, {Soszy{\'n}ski}, {Pietrukowicz},
  {Szyma{\'n}ski}, {Skowron}, {Skowron}, {Mr{\'o}z}, {Poleski}, {Koz{\l}owski},
  {Ulaczyk}, {Rybicki}, {Iwanek}, \& {Wrona}}]{udalski2018}
{Udalski}, A., {Soszy{\'n}ski}, I., {Pietrukowicz}, P., {et~al.} 2018, \actaa,
  68, 315

\bibitem[{{Vall{\'e}e}(2017{\natexlab{a}})}]{vallee2017a}
{Vall{\'e}e}, J.~P. 2017{\natexlab{a}}, The Astronomical Review, 13, 113

\bibitem[{{Vall{\'e}e}(2017{\natexlab{b}})}]{vallee2017b}
{Vall{\'e}e}, J.~P. 2017{\natexlab{b}}, \nar, 79, 49

\bibitem[{{VERA Collaboration} {et~al.}(2020){VERA Collaboration}, {Hirota},
  {Nagayama}, {Honma}, {Adachi}, {Burns}, {Chibueze}, {Choi}, {Hachisuka},
  {Hada}, {Hagiwara}, {Hamada}, {Hand a}, {Hashimoto}, {Hirano}, {Hirata},
  {Ichikawa}, {Imai}, {Inenaga}, {Ishikawa}, {Jike}, {Kameya}, {Kaseda}, {Kim},
  {Kim}, {Kim}, {Kobayashi}, {Kono}, {Kurayama}, {Matsuno}, {Morita}, {Motogi},
  {Murase}, {Nakagawa}, {Nakanishi}, {Niinuma}, {Nishi}, {Oh}, {Omodaka},
  {Oyadomari}, {Oyama}, {Sakai}, {Sakai}, {Sawada-Satoh}, {Shibata},
  {Shizugami}, {Sudo}, {Sugiyama}, {Sunada}, {Suzuki}, {Takahashi}, {Tamura},
  {Tazaki}, {Ueno}, {Uno}, {Urago}, {Wada}, {Wu}, {Yamashita}, {Yamashita},
  {Yamauchi}, \& {Yuda}}]{vera2020}
{VERA Collaboration}, {Hirota}, T., {Nagayama}, T., {et~al.} 2020, \pasj, 72,
  50

\bibitem[{{Vernet} {et~al.}(2011){Vernet}, {Dekker}, {D'Odorico}, {Kaper},
  {Kjaergaard}, {Hammer}, {Randich}, {Zerbi}, {Groot}, {Hjorth}, {Guinouard},
  {Navarro}, {Adolfse}, {Albers}, {Amans}, {Andersen}, {Andersen}, {Binetruy},
  {Bristow}, {Castillo}, {Chemla}, {Christensen}, {Conconi}, {Conzelmann},
  {Dam}, {de Caprio}, {de Ugarte Postigo}, {Delabre}, {di Marcantonio},
  {Downing}, {Elswijk}, {Finger}, {Fischer}, {Flores}, {Fran{\c{c}}ois},
  {Goldoni}, {Guglielmi}, {Haigron}, {Hanenburg}, {Hendriks}, {Horrobin},
  {Horville}, {Jessen}, {Kerber}, {Kern}, {Kiekebusch}, {Kleszcz}, {Klougart},
  {Kragt}, {Larsen}, {Lizon}, {Lucuix}, {Mainieri}, {Manuputy}, {Martayan},
  {Mason}, {Mazzoleni}, {Michaelsen}, {Modigliani}, {Moehler}, {M{\o}ller},
  {Norup S{\o}rensen}, {N{\o}rregaard}, {P{\'e}roux}, {Patat}, {Pena}, {Pragt},
  {Reinero}, {Rigal}, {Riva}, {Roelfsema}, {Royer}, {Sacco}, {Santin},
  {Schoenmaker}, {Spano}, {Sweers}, {Ter Horst}, {Tintori}, {Tromp}, {van
  Dael}, {van der Vliet}, {Venema}, {Vidali}, {Vinther}, {Vola}, {Winters},
  {Wistisen}, {Wulterkens}, \& {Zacchei}}]{vernet2011}
{Vernet}, J., {Dekker}, H., {D'Odorico}, S., {et~al.} 2011, \aap, 536, A105

\bibitem[{{Xu} {et~al.}(2018{\natexlab{a}}){Xu}, {Bian}, {Reid}, {Li}, {Zhang},
  {Yan}, {Dame}, {Menten}, {He}, {Liao}, \& {Tang}}]{xu2018a}
{Xu}, Y., {Bian}, S.~B., {Reid}, M.~J., {et~al.} 2018{\natexlab{a}}, \aap, 616,
  L15

\bibitem[{{Xu} {et~al.}(2018{\natexlab{b}}){Xu}, {Hou}, \& {Wu}}]{xu2018}
{Xu}, Y., {Hou}, L.-G., \& {Wu}, Y.-W. 2018{\natexlab{b}}, Research in
  Astronomy and Astrophysics, 18, 146

\bibitem[{{Xu} {et~al.}(2016){Xu}, {Reid}, {Dame}, {Menten}, {Sakai}, {Li},
  {Brunthaler}, {Moscadelli}, {Zhang}, \& {Zheng}}]{xu2016}
{Xu}, Y., {Reid}, M., {Dame}, T., {et~al.} 2016, Science Advances, 2, e1600878

\bibitem[{{Zhang} {et~al.}(2019){Zhang}, {Reid}, {Zhang}, {Wu}, {Hu}, {Sakai},
  {Menten}, {Zheng}, {Brunthaler}, {Dame}, \& {Xu}}]{zhang2019}
{Zhang}, B., {Reid}, M.~J., {Zhang}, L., {et~al.} 2019, \aj, 157, 200

\end{thebibliography}

\begin{appendix}

\section{The log-periodic spiral model}\label{ap:log_per_spiral}

\begin{figure}[ht]
\includegraphics[width=\columnwidth]{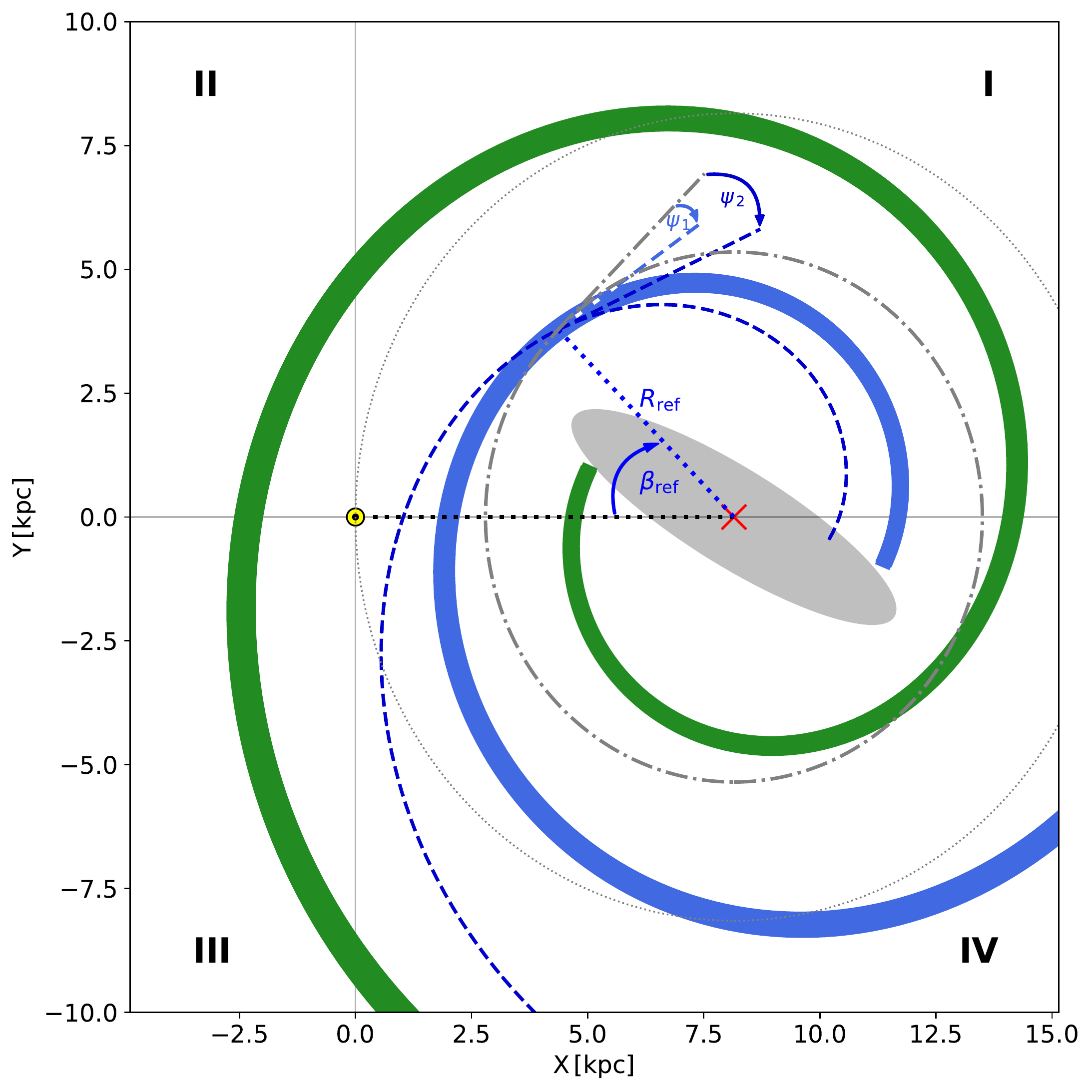}
\caption{Schematic representation of log-periodic spiral arms and the main parameters used. The thick blue spiral has a pitch angle $\psi_1 = 10\degr$. As an example of the effect that an increase of an arm's pitch angle has, while leaving the other 2 parameters fixed, we plotted a log-periodic spiral with $\psi_2=2\times\psi_1$ (blue dashed line). A circle is obtained if $\psi=0$, as indicated by the grey dot-dashed line. As indicated in the Figure, the pitch angle is the angle between the tangent to the spiral arm and a perfect circle, providing a measure of how tightly wound the spiral arm is. The thick green spiral has the same pitch angle as the blue one, but a larger reference radius, $R_{\mathrm{ref}}$. The four Galactic quadrants are indicated by roman numbers I-IV at the corners of the plot. The azimuthal angle, $\beta$, is measured starting from the Sun-Galactic Center direction, and increasing in the direction of Galactic rotation, as indicated here.}
\label{fig:log-periodic_spiral}
\end{figure}

Figure~\ref{fig:log-periodic_spiral} shows the parameters of the log-periodic spiral form that was used to fit the CC positions (see Eq.~\ref{eq:1}). The Galactic quadrants are also indicated there, corresponding, from the first to the fourth quadrant, to the longitude ranges $l \in [0\degr,90\degr]$, $[90\degr,180\degr]$, $[180\degr,270\degr]$ and $[270\degr,360\degr]$.

\end{appendix}

\end{document}